\documentclass[11pt,a4paper]{article}
\usepackage{amssymb}
\usepackage{amsmath}
\usepackage{graphicx}
\begin{document}
{\LARGE{\textbf{India's tryst with modern astronomy}}} \\ \\
Research in astronomy within India has increased substantially over the last two decades, catapulted by growing public awareness about the subject, and efforts of a number of amateur astronomy groups in several towns and cities. The International committee of the Royal Astronomical Society endeavours to promote cohesion amongst its non-UK members, as part of which, several overseas members, including myself, were brought together at a lunch during the recently concluded National Astronomy Meeting. This brief article is a synopsis of my dispensation, intended to provide a glimpse of recent developments in Indian astronomy, at this lunch. Research and public outreach in astronomy, as with all other branches of science in India, is funded by the Department of Science \& Technology (DST) of the Government of India. Public funded science planetariums, generally located in cities, usually arrange a number of events which include documentary films on astronomy, public lectures and open question-answer sessions with astronomers. Many groups of amateur astronomers, accompanied with professional astronomers also visit some of the smaller towns of the country. As a result, the number of young under-graduates wishing to pursue astronomy as their career has increased drastically in the recent past. However, despite the rapidly growing numbers of aspirants, research in this subject is limited to only a handful of specialised institutes. \\ \\
\textbf{Research institutes }
                     India follows a unique model of research where the onus of research lies on institutes outside the purview of local universities, the latter therefore, play a secondary role restricted to taught post graduate courses. Listed below are these institutes: Tata Institute of Fundamental research (TIFR), and its sister institute, National Centre for Radio Astronomy (NCRA), located respectively in Mumbai and Pune; Indian Institute of Astrophysics (IIA), Indian Institute of Science (IISc), and the Raman  Research Institute (RRI), all located in Bangalore; other well-known institutes include the Inter University Centre for Astronomy \& Astrophysics (IUCAA) in Pune, Aryabhatta Research Institute of Observational sciences (AIRES) in Nainital; and the Physical research laboratory (PRL) in Ahmedabad. A new addition to this list is the Indian Institute of Science Education and Research (IISER) located in five cities across the nation. \\ \\                      
\textbf{Major observational facilities } 
                    The Giant Metre wave Radio Telescope (GMRT, Pune)  [50-1420 MHz], and another radio telescope in the southern hills of Nilgiri at Ooty [326 MHz], both operated by the NCRA, along with the one at Gauribidanur [34.5 MHz], built and operated by the IIA are three facilities for radio astronomy in the country. The IIA also owes three more observatories which includes the one in southern hills at Kodaikanal, perhaps one of the oldest in India established in 1899, along with the Vainu Bappu telescope at Kavalur, both of which have telescopes with aperture diameters between 40 inches and 2.3 metres. Two other IIA observatories are located in the Himalayas, viz. the 2 metre Himalayan Chandra telescope (HCT) operated in infrared wavebands, and the High energy gamma ray telescope (HAGAR) capable of detecting 250 GeV photons. This latter telescope is being set-up in collaboration with the TIFR. Recently a 2.3 metre optical telescope was installed in the Himalayan foothills by the AIRES. The Solar observatory at Udaipur, in the North western plains of India, has a number of telescopes with aperture diameters between 15 cm – 25 cm,  is one of the participants in a global network of observatories, GONG, that studies Solar oscillations. \\ \\
\textbf{Principle areas of research }
                             Astronomers at various research institutes listed above extensively use these observational facilities to study active galactic nuclei and the kinematic and physical properties of associated jets, along with other transient objects. Another particularly active area research is the spectroscopic  study of stars and clusters of stars to determine chemical abundances, and physical properties of stellar clusters. Other important developments in the recent past includes the enrollment of India in the Thirty Metre Telescope programme and the development of the space based UV camera (UVIT),  built in collaboration with the Herzberg Research Institute, Canada, operated in the 1200 – 3000 Angstrom waveband with a resolution of 1.5 arc second. 

Violent phenomena on the Solar surface, such as massive eruptions of Solar mass called coronal mass ejections (CMEs) and the associated radio emissions, called Noise storms, are studied extensively using various radio telescopes listed above.  The Solar observatory at Udaipur is primarily used to study features of the Solar disk and radial oscillations of the Sun.
It is evident that astronomers in India are making rapid forays in numerous areas of the subject, and with a number of observational facilities around the country it could well be an attractive destination for others across the globe. Further details may be found on websites of individual research institutes.
\end{document}